\documentclass{IEEEtran}
\IEEEoverridecommandlockouts
\usepackage{amsmath, amssymb, amsfonts}
\usepackage{algorithmic, graphicx, textcomp, xcolor}
\usepackage{tabularx}
\usepackage{array}
\usepackage{xcolor}
\usepackage{fancyhdr}
\newcolumntype{S}{>{\hsize=.5\hsize\linewidth=\hsize}X}
\newcolumntype{L}{>{\hsize=1.5\hsize\linewidth=\hsize}X}    %the sum of all columns widths must add up to a multiple of X

\usepackage[
style=ieee,doi=false,isbn=false,url=false,eprint=false,maxnames=4
]{biblatex}
\begin{document}

\title{~\vspace{-12mm}\\{\normalsize preprint to appear in IEEE Transactions on Power Systems, accepted December 2023. DOI 10.1109/TPWRS.2023.3342729\\
\vspace{-5mm}
\copyright 2023 The Authors. This work is licensed under a Creative Commons Attribution 4.0 License.}\\
\vspace{1mm} Towards using utility data to quantify how investments would have increased the wind resilience of distribution systems
%Measuring increased resilience from wind resilience investment using distribution utility outage data\\
}

\author{Arslan Ahmad,~\IEEEmembership{Student Member~IEEE} \qquad\qquad Ian Dobson,~\IEEEmembership{Fellow~IEEE}
\thanks{A. Ahmad and I. Dobson are with the Electrical and Computer Engineering Department,
Iowa State University,
Ames, IA, USA. 
dobson@iastate.edu. Funding from the U.S. Government, the U.S. Department of State, IIE, and USEFP via the Fulbright program and USA NSF grants 2153163 and 2228757 is gratefully acknowledged.}
}

%\fancyhead[c]{\textnormal{ preprint to appear in IEEE Transactions on Power Systems, accepted December 2023. doi: 10.1109/TPWRSxxxxx}}
%\renewcommand{\headrulewidth}{0.0pt}
%\fancyfoot[C]{\fontfamily{ptm}\selectfont\fontsize{10}{10}This work is licensed under a Creative Commons Attribution 4.0 License; see http://creativecommons.org/licenses/by/4.0/}
%\fancyfoot[c]{\textnormal{\small This work is licensed under a Creative Commons Attribution 4.0 License; see http://creativecommons.org/licenses/by/4.0/}}
% \fancyfoot[L]{~\\[-20pt]Preprint \copyright 2016 IEEE. \small Personal use of this material is permitted. Permission from IEEE must be obtained for all other uses, in any current or future media, including reprinting/republishing this material for advertising or promotional purposes, creating new collective works, for resale or redistribution to servers or lists, or reuse of any copyrighted component of this work in other works.}
%  \fancyfoot[C]{~ }

\maketitle

\begin{abstract}
We quantify resilience with metrics extracted from the historical outage data that is routinely recorded by many distribution utilities.
The outage data is coordinated with wind data to relate average outage rates in an area to wind speed measured at a nearby weather station. 
A past investment in wind hardening would have reduced the outage rates, and the effect of this on metrics can be calculated by sampling a reduced number of historical outages and recomputing the metrics. 
This quantifies the impact that the hardening would have had on customers. 
This is a tangible way to relate an investment in wind resilience to the benefits it would have had on the lived experience of customers that could help make the case for the investment to the public and regulators.
We also quantify the impact of earlier or faster restoration on customer metrics and compare this to the impact of investment in hardening. Overall, this is a new and straightforward approach to quantify resilience and justify resilience investments to stakeholders that is directly driven by utility data. The approach driven by data avoids complicated models or modeling assumptions.
\end{abstract}


%ISSUES

\section{Introduction}
Resilience generally addresses the response of power systems to extreme weather events, as well as other unusual high stresses such as earthquakes, fires, and epidemics. 
There are many overlapping frameworks and definitions of resilience \cite{StankovicPS22,PanteliPROCIEEE17,NanRESS17, ciapessoniCIGRE23} and the most concise definition is ``Power system resilience is the ability to limit the extent, severity, and duration of system degradation following an extreme event." \cite{CIGREELECTRA19}. There remains much scope for practically quantifying resilience \cite{JiPROCIEEE17} and for justifying investments that improve resilience. 
In particular, this paper proposes a data-driven approach to quantify the resilience of a distribution system to wind\footnote{Regarding the definition of wind resilience, while some of the nuances of the overall definition of resilience are vigorously debated, e.g., \cite{ciapessoniCIGRE23}, distinguishing types of resilience by threat, such as wind, seems widely accepted. 
Some other resilience threats may, to some extent, involve wind (e.g., ice storms, extreme rain, wildfires) or may be independent of wind (e.g., cyber attacks, physical attacks, earthquakes, geomagnetic disturbances).} and to help justify investments in hardening the system or speeding up its restoration. Wind-related hazards due to storms and hurricanes are one of the most significant hazards for overhead distribution system; the effects of high winds include tree falls and flying debris as well as direct impacts such as pole toppling and conductor galloping.

Investment decisions upgrading distribution systems or their restoration consider many factors, such as cost, reliability, load, distributed generation, deployment of crews and materials, and many engineering and community constraints. 
In order to also consider resilience in these decisions, it is desirable to quantify the benefits of investments that increase resilience and communicate those benefits to utilities, communities, and regulators. 
One useful but complex approach is to make detailed models of the extreme weather, its impact on the distribution system, the restoration process, and the impact on customers and then use these models to estimate the future benefits of a proposed upgrade. 
This model-based approach is challenging because of the extensive approximations and assumptions needed to make practical models of the entire process. Nevertheless, there has been considerable progress in using these models as reviewed in section \ref{litreview}. 
The purpose of this paper is to open up another, complementary approach driven directly by utility data that can also help to quantify and communicate the benefits of investments that increase resilience to strong winds. 

\begin{figure}[h]
    \centering
    \includegraphics[width=0.48\textwidth]{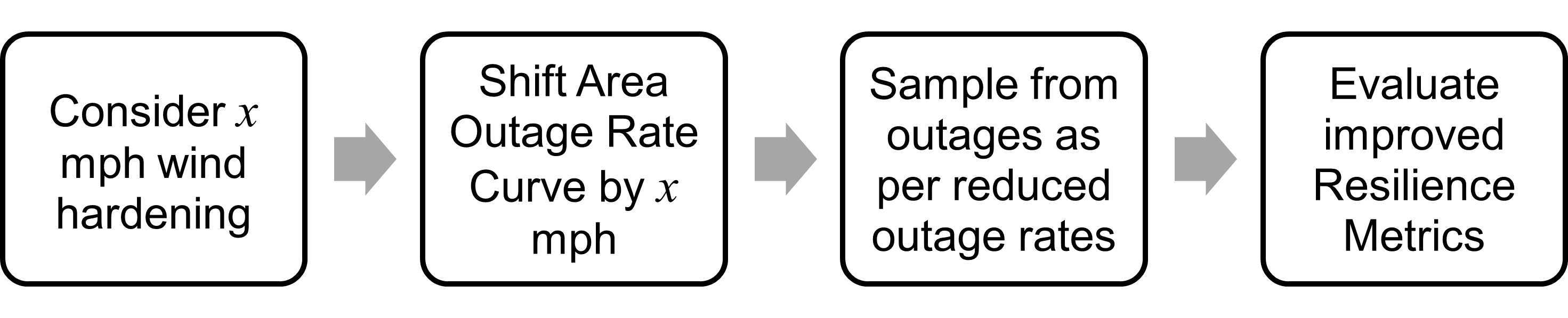}
    \caption{Quantifying the benefits of wind-hardening resilience investments.}
    \label{fig:flowDiagram}
\end{figure}

We briefly outline the new data-driven approach in Fig.~\ref{fig:flowDiagram} and as follows:
Consider an area inside the distribution system that is close to a weather station measuring wind speed. 
We process the area outage data and the wind speed together to obtain an ``area outage rate curve" that describes how the mean outage rate of the area increases with wind speed. 
The area outage rate curve quantifies the resilience of the area with respect to the measured wind speed. 
A previous investment in hardening the area would have shifted the area outage rate curve and reduced the area outage rates.
We can go back to the historical outage data and sample a reduced number of outages from it to match the reduced outage rates. 
This samples the historical outages that the investment would have caused. Computing resilience metrics for the historical outages and comparing these to the improved resilience metrics for the sampled outages quantifies the effect that the investment would have had on the area and its customers.
Moreover, we also quantify the impact of improved restoration:
Investment in restoration can make the restoration start earlier or increase the rate of restoration.
We change the historical restoration times to correspond to these improvements and recalculate the metrics to quantify their impacts.
This brief outline is elaborated throughout the paper, but we start by discussing the value of the new approach in justifying resilience investments.

In addition to the routine reluctance to pay more for improved electrical infrastructure, resilience investments are particularly hard to justify because they concern rarer large events that will recur, but at an indefinite time in the future\footnote{Note that reliability investments already address the common failures.}. 
There are indeed technical difficulties in estimating the impact of rare events, but arguably at least as important for practically improving resilience are the difficulties in communicating the impact of rare events and the benefits of investing to reduce their impact. 
The model-based approach can estimate, with some approximations and assumptions, the projected future benefits of a resilience investment. The data-driven approach can estimate the benefits that the investment would have had if the investment had previously been made. 
%In some ways, the data-driven approach can be more persuasively tangible because it is related to people's lived past experience. 
In some ways, the approach driven by historical data can be more persuasively tangible because the historical data is related to people’s lived past experience, in that the historical blackout risk and consequences are more easily recalled and more vivid than imagining a blackout predicted by a model at an indeterminate time in the future.
For example, it could be persuasive to say that the community would have saved 20\% fewer customer minutes of outages over the past 10 years, or even more specifically that a particular, extreme wind event (such as the upper midwest USA derecho in August 2020 that caused $\sim$11 billion dollars of damage) would have had 20\% fewer customer minutes out.

The overall paper contribution is 
showing the feasibility of a new, entirely data-driven method of quantifying distribution system resilience to wind, particularly the change in standard metrics that would have occurred if overall investments in hardening or earlier or faster restoration had been made. 
There are no modeling assumptions,
the data is readily available to utilities with an outage management system,
and the computations are fast and relatively straightforward.
The new quantification of resilience benefits of investments can be tangible to stakeholders and help support the case for investments.
The more detailed technical contributions of the paper include implementing the overall effect of hardening by sampling a reduced number of historical outages and the new conception and use of area outage rate curves and super events.

After reviewing the literature in section~\ref{litreview}, section~\ref{DataSection} 
describes the outage and weather data used. Section \ref{outageratecurves} describes 
the area outage rate curves 
that quantify the wind resilience from data, and how they can be shifted to represent hardening.
Section \ref{MetricsSection} explains how to extract resilience events from data and calculate their metrics.
Section \ref{SamplingSection} describes the sampling that represents the hardening and 
Section \ref{restoration} shows how improved restoration is represented.
Results quantifying the improvement in metrics due to the hardening and improved restoration 
are presented in Section \ref{results}.
Sections \ref{outageratecurveDerivationSection} and \ref{superevents} give technical details of constructing area outage curves and tracking events after sampling.
Finally, the paper contributions and conclusions are given in section \ref{conclusions}.
The paper elaborates and builds on the initial work presented in the MS thesis \cite{AhmadMS23}.

\section{Literature review}
\label{litreview}

Since high winds are a significant hazard for overground power distribution systems \cite{AhmadMS23,climate22}, there is substantial work studying the resilience of these systems to the wind.

We begin by briefly surveying a variety of methods that build and use models to quantify distribution system resilience.
These methods form a useful complementary approach to our new data-driven quantification of resilience.
Xu \cite{XuQuanta08} defines a calculator for hurricanes
that calculates the cost and benefits of distribution system undergrounding or hardening given 
utility estimates of input data. The modeling includes exponential fragility models for poles, power law fragility models for span damage, 
simulated hurricanes, and restoration times estimated from crew availability.
Ma \cite{MaPS18} formulates system hardening, the impacts of extreme weather, and minimizing load shedding as a multi-level mixed-integer linear program to find an optimal hardening 
strategy. Arif \cite{ArifPS18} co-optimizes distribution system operation and repair crew routing for outage restoration after extreme weather events using a two-stage stochastic mixed integer linear program.
%Ma \cite{MaPESGM19} extends the optimization to include modeling of all phases from preparation and hardening to restoration.
Tan \cite{TanPS18} finds an optimal hardening and repair sequence to minimize the expected energy not served using mixed integer 
linear programming and associated convex relaxations and heuristics.
Tan \cite{TanPS19} finds optimal repair sequencing when there is large-scale damage by solving a scheduling problem 
using approximations to linear programming and accounting for
multiple faults obstructing the power to the same customer.
Ouyang \cite{OuyangSS14} models the resilience of Harris County, Texas, to hurricane Ike using exponential fragility curves, a DC power flow network model, and high-level models of restoration resources and sequencing.
Ciapessoni \cite{CiapessoniENERGIES21} evaluates the power grid's resilience in a mountainous area to the combined effects of wind, snow, and trees with a risk-based model.
Poudel \cite{PoudelSJ19} simulates for a distribution of wind conditions the improvement in resilience risk of planned improvements using measures such as value at risk of energy not served.
Wei \cite{WeiAM16} analyzes resilience to particular severe hurricanes with time-varying Poisson processes that model distribution system outage and repair processes as a queue. 
Hughes \cite{hughesRESS21} develops a wind fragility based weather infrastructure model to predict outages under extreme weather events and then analyzes the cost and benefits of different infrastructure hardening scenarios on the resilience of the overhead distribution system by simulating outages under each scenario using Monte Carlo. 

We now survey methods that analyze dependencies in distribution system data to describe 
wind resilience. Much of the analysis is for specific hurricanes.
Davidson \cite{DavidsonNHR03} studies resilience to hurricanes in the Carolinas using utility outage data, a combination of interpolated and simulated hurricane wind data, and land cover and rain data. 
Liu \cite{LiuRESS08} constructs a spatial generalized linear mixed model of the number of hurricane and ice storm outages in zip codes as a function of weather, protection device density, and spatial data. 
Reed \cite{ReedSJ09} studies the resilience to wind-induced damage in Hurricane Katrina. 
Jaech \cite{JaechPS19} estimates individual component repair times using a gamma distribution with parameters predicted using neural networks trained on utility outage records.
%For example, Jaech \cite{JaechPS19} predicts individual component outage restoration times and customer hours lost with a neural network, and
%Chow \cite{ChowPD96} analyzes the contributions of timing, faults, protection, outage types and weather to individual distribution system component restoration times.
%Some papers use gamma distributions to fit outage times or restoration times \cite{JaechPS19,CarringtonPS21}
Carrington \cite{CarringtonPS21} establishes methods for extracting events from utility outage data, computes standard resilience metrics for the events, and estimates restoration times.
%Related methods for transmission systems are in \cite{EkishevaPMAPS22}.
Cerrai \cite{cerraiACCESS19} and Kezunovic \cite{kezunovicHICSS18} use machine learning methods based on multiple decision trees and logistic regression to combine weather prediction, vegetation, outage, and other data to predict the probability of storm outages in small areas of the distribution system. 
Al Mamun \cite{almamunNAPS23} studies the impacts of extreme weather events on the reliability and resilience of a North Carolina distribution system by analyzing the correlations in outage data between outage duration, the number of affected customers, and SAIDI.

Now we review relevant work extracting fragility curves from wind and outage data for use in resilience models. Dunn \cite{DunnNHR18} develops empirical fragility curves for overhead lines from 11 kV to 132 kV in the UK during wind storms, defined as winds exceeding 38 mph. 
The fragility curves are fit with power laws, and relate the number of faults per line length in areas of $\sim$2000 km$^2$ to the reanalyzed maximum wind gust over the storm duration in the areas.
Reed \cite{ReedWEIA08} analyzes the fragility and outage duration of an urban distribution system for four different wind storms. Fragility curves show the fraction of affected feeder length as a function of the peak storm wind gust squared. 
Bjarnado \cite{BjarnadoIS13} describes lognormal fragility curves for pole design subject to hurricane risk as well as reviewing deterministic pole design with safety factors. 
Murray \cite{MurrayPMAPS14} correlates reanalyzed wind gust data with transmission system faults in the UK to give an exponential fragility curve for transmission lines. 
Hughes \cite{hughesRESS22} calibrates physics-based structural fragility curves of overhead distribution systems using machine learning techniques to compensate for the limited empirical data for large events, and develops a hybrid physics-based and data-driven outage prediction model with improved accuracy.
{Donaldson \cite{donaldsonMA23} uses historical outage data to develop fragility curves normalized to expected wind speeds in each region in order to compare the wind resilience between several regions of a distribution network operator in northwest England.

Reliability addresses performance averaged over the year rather than resilience, but we acknowledge the extensive and useful tradition of evaluating distribution system reliability with steady-state Markov models \cite{BillintonAllanBook} in which extreme weather is modeled with additional Markov states \cite{BillintonGTD06}. 

\section{Outage data and weather data} \label{DataSection}
Two different datasets are used, one with outage data recorded by a distribution utility in the USA and the other containing NOAA (National Oceanic and Atmospheric Administration) wind data for weather stations in that utility's service area.
Both datasets cover the same time span of six years.
The outage dataset contains 32\,278 individual outages and has a one-minute resolution; i.e., all outages that occurred within one minute are labeled with the same timestamp.
Each outage entry in the dataset corresponds to an outage of a component in the power distribution system and includes the location coordinates of that component, the number of customers affected during the outage, the outage's starting and ending time, and cause codes.\footnote{149 outage records with missing location information are removed. Further data cleaning and pre-processing details are given in \cite{AhmadMS23}.}
%The cause codes are assigned to each outage by the utility to indicate the reason behind it, such as weather, trees, equipment failure, and accident.

Each record in the weather dataset provides the average hourly wind speed and various other weather details. 
%of different observed weather parameters, including wind speed, wind gust speed, wind direction, precipitation, sky conditions, relative humidity, and monthly summaries; however, 
Only the average hourly wind speed data is used here.
There is data from multiple weather stations in the utility's service area.
With the outage and weather stations' geographical locations known, each outage is associated with the weather station closest to the outage. 
This divides the distribution system into multiple areas, where each area contains all the outages closest to the weather station associated with that area. This paper analyzes the outages in two of these areas, as shown in Fig.~\ref{fig:outageLocations}.
Area 1 and area 2 contain 7876 and 12\,715 outages, respectively.

The typical measured wind speeds at the weather station in area 2 are systematically slower than the typical measured wind speeds at the weather station for area 1. Differences in these two measurements are expected: 
Weather Station 2 measures wind 5 feet above the ground in a field surrounded by woods, whereas Weather Station 1 measures wind 33 feet above the ground in a large, flat area with no trees.
The weather stations are 13 miles apart.

When applying the method of this paper in practice, we note that the data should be collected over a period of time long enough to have sufficient large events with high wind (in our case, 6 years of data gave 32 large events in Area 1 and 88 large events in Area 2) and that it is better to use data over a recent time period. 

\begin{figure}[ht]
    \centering
    \includegraphics[width=0.48\textwidth]{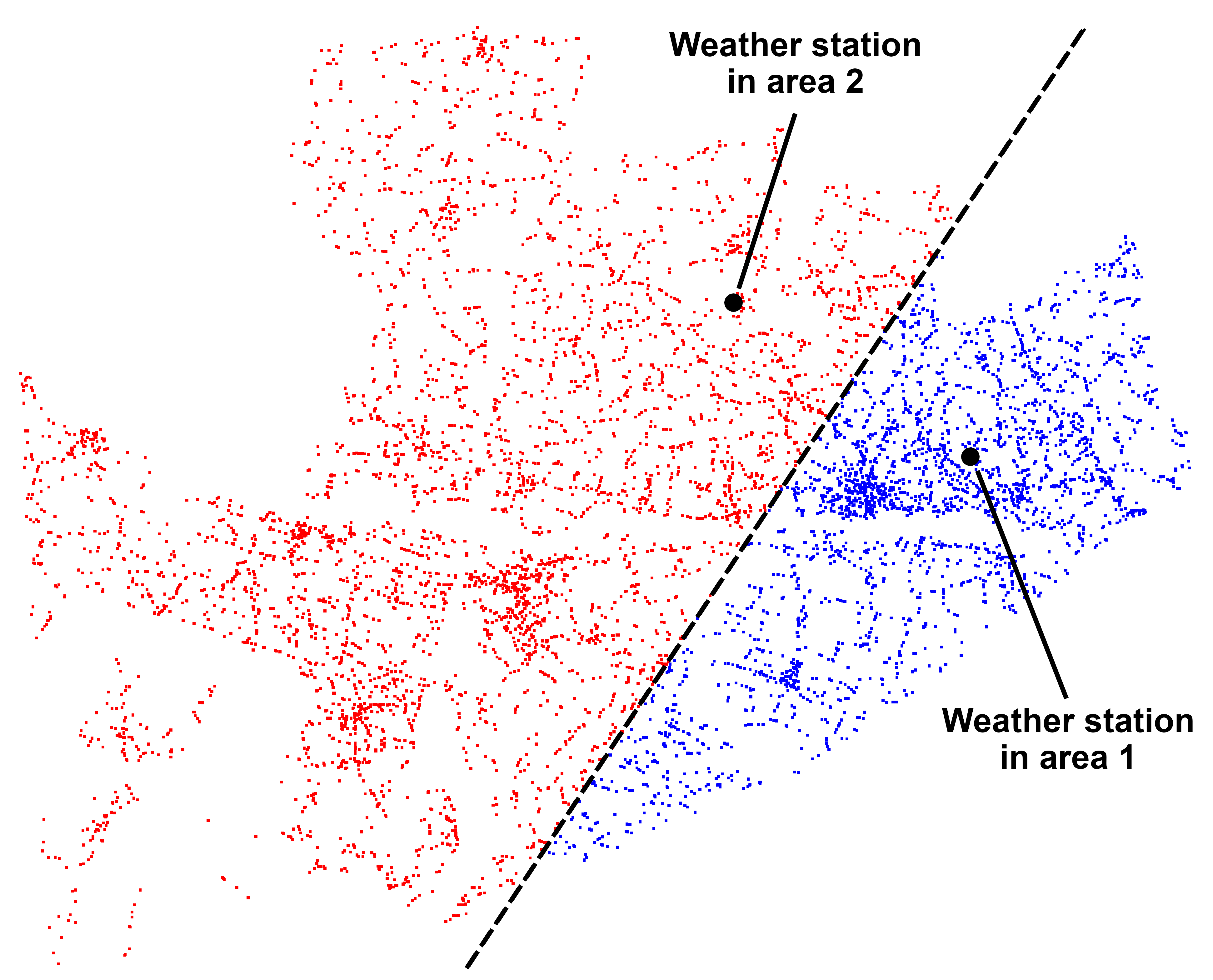}
    \caption{Geographical location of outages and the associated weather stations in two areas of a distribution system.}
    \label{fig:outageLocations}
\end{figure}

\section{Area outage rate curves}
\label{outageratecurves}

Area outage rate curves describe the resilience to wind of an area of the distribution system based on the average outage rates observed at different wind speeds. This section explains area outage rate curves, how they quantify wind resilience, and how shifting them represents the 
effect of distribution system hardening.
%Section \ref{outageratecurveDerivationSection} gives the details of calculating area outage rate curves from the outage and wind data.

\subsection{Quantifying wind resilience with area outage rate curves}

Each area of the distribution system is associated with its weather station.
The area outage rate curve specifies the mean outage rate of the area $\overline{F}(v)$ as a function of the wind speed $v$ measured at the weather station.
In Figs. \ref{fig:MBExponentialModel} and \ref{fig:PKExponentialModel},
the dots indicate the empirical mean outage rates  $\overline{F}^{\rm empirical}(\hat v)$ at integer wind speeds $\hat v$ that are calculated from the wind and outage data for areas 1 and 2; the details of the calculation are given in section \ref{outageratecurveDerivationSection}. Whereas the area outage rate curves in Figs. \ref{fig:MBExponentialModel} and \ref{fig:PKExponentialModel} are exponential fits to the empirical data of the form
\begin{align}
\label{eq:exponentialEquation}
   \overline{F}(v)=ae^{bv}
\end{align}
The exponential fit\footnote{The exponential has simplicity and a better fit to our data than alternatives such as power law and exponential fit with offset. The fit uses the NonLinearModelFit function in Mathematica that implements the Levenberg-Marquardt method.} uses the Levenberg-Marquardt method (also known as the damped least-squares method) with a 99\% confidence level for parameters and predictions. The parameter values obtained for the fit are $a=0.00018$ and $b=0.38$ for area 1, and $a=0.006$ and $b=0.48$ for area 2.
The wind station in area 2 measures slower wind speeds than the wind station in area 1. This is expected as discussed in section \ref{DataSection}.

\begin{figure}[ht]
    \centering
    \includegraphics[width=0.48\textwidth]{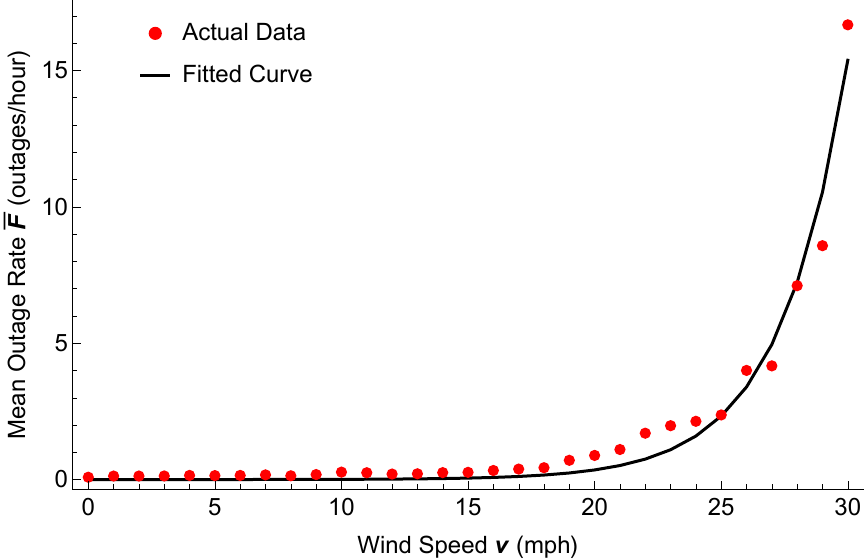}
    \caption{Area outage rate curve of area 1. Dots are the mean outage rate at each wind speed from data, and the curve is an exponential fit.}
    \label{fig:MBExponentialModel}
\end{figure}
\begin{figure}[ht]
    \centering
    \includegraphics[width=0.48\textwidth]{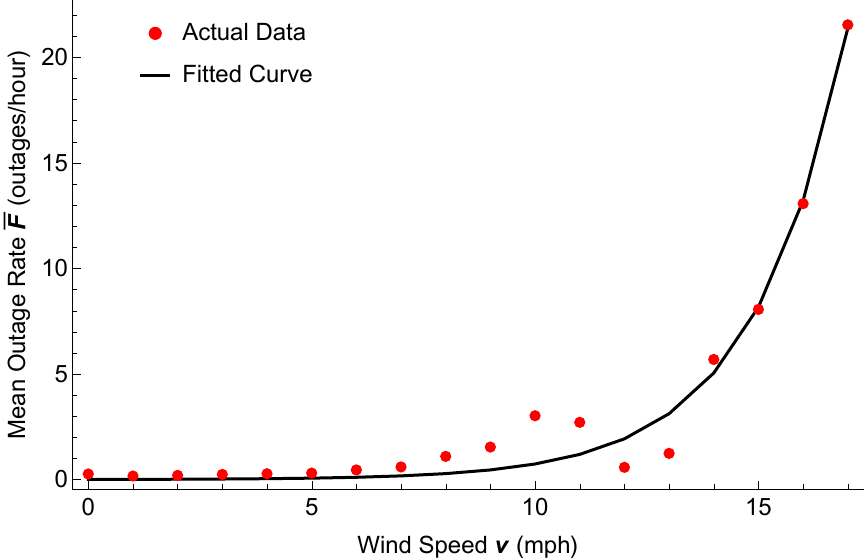}
    \caption{Area outage rate curve of area 2. Dots are the mean outage rate at each wind speed from data, and the curve is an exponential fit.}
    \label{fig:PKExponentialModel}
\end{figure}

The outage rate curve describes the resilience of the area with respect to the wind measurement at the weather station.
The mean outage rate is low, except that it increases sharply for higher wind speeds. The sharp increase in mean outage rate is expected: As the wind speed increases, the overhead distribution infrastructure (poles, wires, cross arms, insulators) and nearby vegetation face more stress, leading to more faults and increased outage rates. Other authors report a dependence of outage rate data on the wind speed of a similar form and use exponential \cite{BrownQuanta09,XuQuanta08,MurrayPMAPS14} or power law \cite{DunnNHR18,XuQuanta08} fits to describe this dependence.

\subsection{Hardening shifts the area outage rate curves}
\label{hardening}

Overhead power line components such as poles are designed to withstand their rated wind speed. Hardening upgrades or reinforces the components. 
The overall effect of the hardening is that the same mean outage rate can be achieved at a higher wind speed; that is, the hardening shifts the outage rate curve to the right, as shown in 
Fig.~\ref{fig:shiftedCurvesOverallMB}.\footnote{A right-shift is also used to model the failure rate data for hardened transmission structures in Richard Brown's report \cite[Figs. 5-4 and 5-5]{BrownQuanta09}.} For example, replacing poles rated for 60 mph wind with poles rated at $(60+x)$ mph for a positive $x$ value would shift the outage rate curve $\overline{F}(v)$ right by $x$ mph so that the new outage rate curve is $ \overline{F}_{new}(v)=\overline{F}(v-x)$. 
Since the mean outage rate generally increases with wind speed, $\overline{F}_{new}(v)<\overline{F}(v)$ so that the hardening reduces the mean outage rate at wind speed $v$. 
This reduction in the outage rate is implemented with sampling in section~\ref{SamplingSection}. In the case of the exponential area outage rate curve (\ref{eq:exponentialEquation}), the reduction in outage rate takes the simple form of multiplying the outage rate by the same factor $e^{-bx}$ at all wind speeds:
\begin{align}
   \overline{F}_{new}(v)=\overline{F}(v-x)=ae^{b(v-x)} = \overline{F}(v)e^{-bx}  \label{eq:exponentialModelShift}
\end{align}
We also note that for the exponential outage rate curve, the right shift is equivalent to the alternative modeling of reducing all the outage rates by the same factor.

\begin{figure}[ht]
    \centering
    \includegraphics[width=0.48\textwidth]{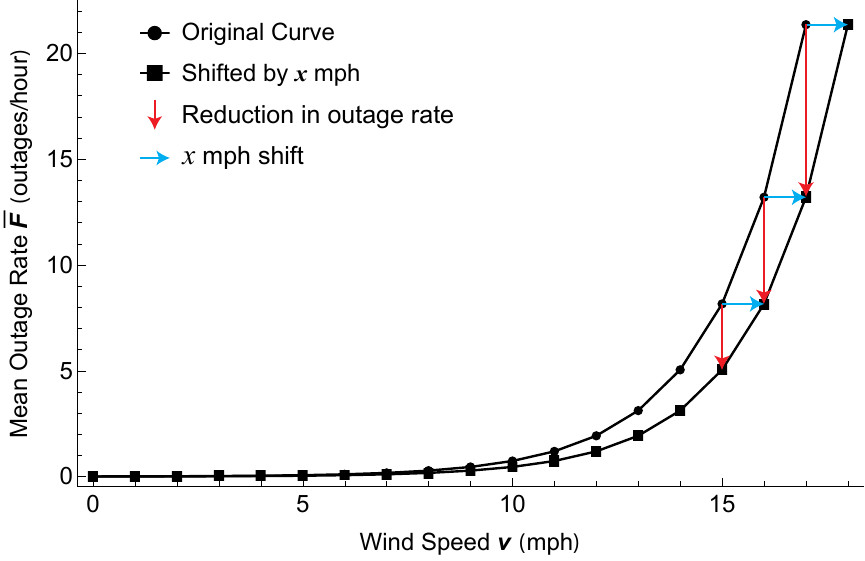}
    \caption{Comparison of original and shifted area outage rate curve of area 2 after 1 mph hardening to wind hazard.}
    \label{fig:shiftedCurvesOverallMB}
\end{figure}

\subsection{Comparing area outage rate curves and fragility curves}

Fragility curves for wind describe the probability of component failure or failure per kilometer of line as a function of wind speed \cite{DunnNHR18, MurrayPMAPS14,ReedWEIA08}.
Area outage rate curves have some similarities and differences with fragility curves, so it is interesting to compare them. 
Area outage rate curves give the mean outage rate of an area as a function of the measurement of wind speed at a particular weather station, whereas fragility curves give the probability of failure of a component as a function of the wind speed (at least conceptually) at that component. The wind speed at any given component in the area is correlated with, but different than the wind speed at any particular weather station.
%\footnote{For example, the correlation between wind measured at two nearby points is used when siting wind turbines to relate measurements for a limited time at the proposed site with many years of measurement at a weather station cite MPE methods.}.
The area contains many types of components that can cause an outage (poles, lines, insulators, etc.), and these components can also differ in manufacturer, age, elevation, topography, local environment, tree cover, and condition and are subject to different wind conditions. 
Since the area outage rate curve is directly obtained from historical data, it incorporates all these types and variations, and describes the aggregated response of the area in terms of outages with respect to the wind speed at the particular weather station. 
One notable difference is in the use of the two curves: area outage rate curves bypass any component modeling to directly describe the aggregated resilience of the entire distribution system area with respect to a particular wind measurement, whereas fragility curves are used in models of components to design that component or to compute the resilience of many similar components. Indeed, it is not appropriate to directly substitute an area outage rate curve for a component fragility curve in models.

%The mean outage rate is related to the failure probability under some assumptions: if one models the distribution system area as $N$ similar components under wind stress that is constant over a time period and assumes independent component failures, then the mean outage rate of the area is proportional to $N$ times the individual component failure probability.
%One aspect is that an area outage rate curve involves time differently than a fragility curve: A fragility curve gives a probability of failure if a given wind speed is applied to the component without considering the time at or over which the wind speed is applied, whereas an outage rate inherently applies over a time interval. To address this, fragility curves can be formulated as depending on the maximum wind speed over an event cite Dunn.
%To enable use in models and predictions, fragility curves can be derived from data to give failure probabilities per km of line (\cite{DunnNHR18, DunnICASP2015})cite reed.

%The area outage rate curves are overall similar to the fragility curves introduced by Dunn et al. (\cite{DunnNHR18, DunnICASP2015}), but there are many detailed differences in the processing and some differences in interpretation/application.
%The fragility curves by Dunn et al. show a relationship between the average number of faults/1000km in a distribution system during wind storm events (defined by wind speeds $\geq$ 17/ms).

\section{Constructing area outage rate curves}
\label{outageratecurveDerivationSection}
This section explains how to align and process the outage and wind data to construct the area outage rate curves. 

Since the outage times are recorded to the nearest minute and the wind speeds are recorded hourly\footnote{Eighteen outages with a time difference of more than 201 minutes and low wind speeds are omitted from further analysis. Some of the wind speed data is at 15-minute or more than one-hour intervals. The time zone convention of each wind station should be checked, since their use of daylight savings time can vary.},
we need to interpolate the wind speeds.
Let $V(t)$ be the piecewise linear interpolation of the wind data as shown in Fig.~\ref{fig:windSpeedInterpolation}.

\begin{figure}[ht]
    \centering
    \includegraphics[width=0.48\textwidth]{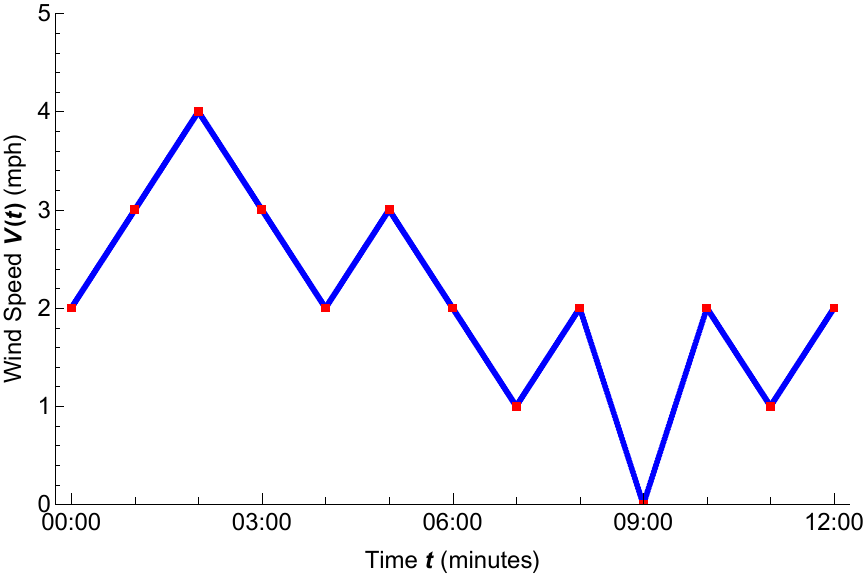}
    \caption{Piecewise linear interpolation $V(t)$ of hourly wind speed data. Red dots represent the wind speeds at each hour and the blue lines represent the interpolated wind speeds.}
    \label{fig:windSpeedInterpolation}
\end{figure}

\begin{figure}[ht]
    \centering
    \includegraphics[width=0.48\textwidth]{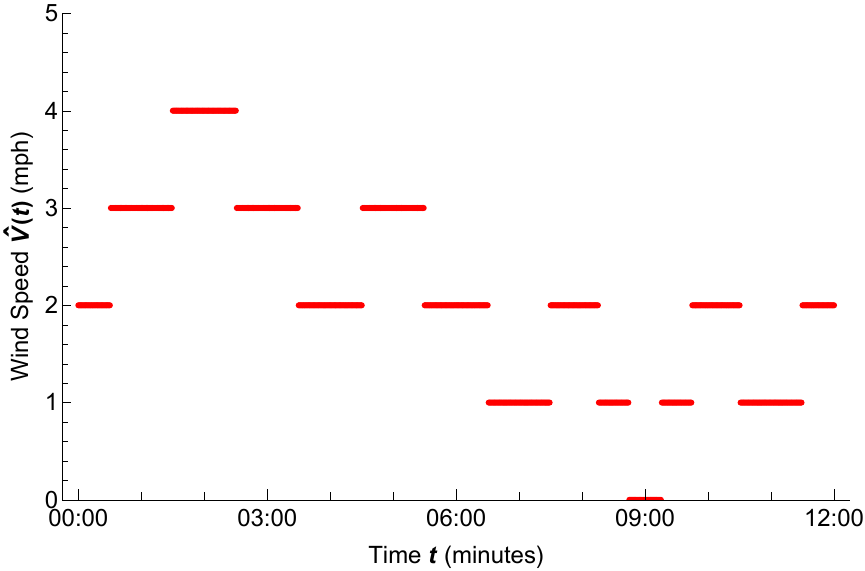}
    \caption{Wind speed $\hat V(t)$ that is rounded to the nearest integer wind speed.}
    \label{fig:windSpeedDiscretization}
\end{figure}

The wind speed as a function of time $t$ can be rounded\footnote{We use statistician's rounding that rounds the borderline cases to the nearest even integer.} to the nearest integer wind speed in miles per hour to obtain the integer wind speed function 
\begin{align}
\hat V(t)=\mbox{round}(V(t))
\end{align}
as shown in Fig.~\ref{fig:windSpeedDiscretization}.
For a given integer wind speed $\hat v$, 
\begin{align}
 \hat V^{-1}(\hat v) =\{ t\mid \hat V(t)=\hat v  \}
\end{align}
is all the times for which the wind speed rounds to $\hat v$. $ \hat V^{-1}(\hat v)$ is a set of time intervals.
The total time in $ \hat V^{-1}(\hat v)$ is the sum of the durations of all the time intervals. 
Then the empirical mean outage rate at integer wind speed $\hat v$ is
\begin{align} 
\label{eq:Fempirical}
    \overline{F}^{\rm empirical}(\hat v)=
    \frac{\mbox{number of outages occurring  in  } \hat V^{-1}(\hat v)}{\mbox{total time in } \hat V^{-1}(\hat v)}
\end{align}
Evaluating $\overline{F}^{\rm empirical}$ using (\ref{eq:Fempirical}) at integer wind speeds $\hat v$ gives the empirical mean outage rates shown as dots in Figs. \ref{fig:MBExponentialModel} and \ref{fig:PKExponentialModel}.

\section{Events, processes and resilience metrics}  \label{MetricsSection}
A necessary step in the data processing groups outages into resilience events and then calculates several metrics for each event.
The particular metrics we use are among the typical resilience metrics proposed and explained in references such as \cite{PanteliPS17FLEP,RaoufiSUSTAIN20}, allowing for the observation that in real data the theoretically successive phases of resilience usually overlap \cite{CarringtonPS21}.
%To do this we apply the methods of \cite{CarringtonPESGM20}. This section summarizes these methods, relying on \cite{CarringtonPESGM20} for a detailed description. 
The focus on events, particularly the larger ones, and the metrics for events instead of average performance over a year make this a resilience analysis. 

To define the resilience events and automatically extract them from the distribution system data we use the method in \cite{CarringtonPESGM20,CarringtonPS21}. %\footnote{Related methods for transmission systems are in \cite{EkishevaPMAPS22}.} 
The start of an event is defined by an initial outage that occurs when all components are operational, and the end of the same event is defined by the first subsequent time when all the components are restored. 
We write $n$ for the number of outages in an event.
If we write $o_1$ for the start time of the first outage and $r_n$ for the time of the last restore, then the event occurs over the time interval $[o_1,r_n]$.
Two example events are shown in Fig.~\ref{fig:events}.

\begin{figure}[h]
    \centering
    \includegraphics[width=0.48\textwidth]{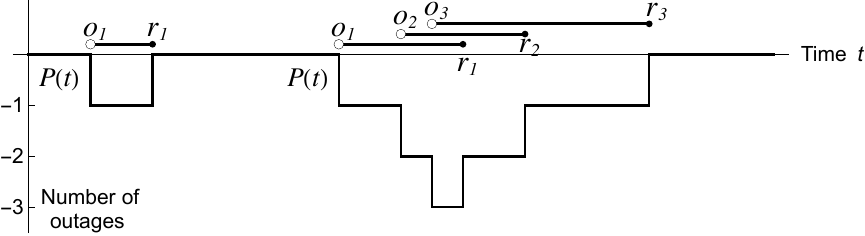}
    \caption{An event with one outage and an event with 3 outages. Above the time axis shows each outage start time (open circle) and restore time (dot). Below the time axis is the performance curve $P(t)$ for each event.}
    \label{fig:events}
\end{figure}

In most of the literature (e.g. \cite{DavidsonNHR03, ReedSJ09, ReedWEIA08, MurrayPMAPS14}), the events in the power system data are identified based on the time period in which the weather is intense, but we use a different approach that identifies events based on the time of occurrence and restores of outages that detects when outages overlap \cite{CarringtonPS21}. 
Our approach based on outages does produce events caused by wind; indeed, almost all large-size events identified in the studied distribution system are caused by high wind speeds. 
An important advantage of our approach is that it allows analysis of events of all sizes and causes. This will be very useful in future work quantifying and comparing the risks of the various causes and sizes of events in distribution systems. 
Events also are defined in transmission systems based on the time of occurrence and restores of outages, and these events are useful in yielding the statistics of North American transmission events of a range of sizes and causes \cite{EkishevaPMAPS22} and their typical stochastic models \cite{DobsonPS23,DobsonPESL23}. 

Performance curves that track in time the negative of the cumulative number of unrestored outages or customers or other quantities are routinely used in studies of resilience to track the progress in time of resilience events \cite{PanteliPS17FLEP,NanRESS17,PanteliPROCIEEE17,StankovicPS22,PoudelSJ19}.
Accordingly, we define the component performance curve $P(t)$ as the negative of the cumulative number of unrestored outages in an event.
($P(t)$ is also the cumulative number of restores at time $t$ minus the cumulative number of outages at time $t$ \cite{CarringtonPS21}.)
Component performance curves for two events are shown in Fig.~\ref{fig:events}.
$P(t)$ decrements by one when there is an outage and increments by one when there is a restore.
In particular, $P(t)$ is initially zero, the event starts at $o_1$ when the cumulative number of failures $P(t)$ first decrements from zero, and ends at $r_n$ when $P(t)$ increases to return to zero. 
The events group together the successive outages that have some overlap in duration. 
Events in our distribution utility data (2944 total events in area 1 and 3706 in area 2) are of all sizes, ranging from a single outage that is restored without involving any other outages to the largest event with more than 100 overlapping outages. 

In an event with $n$ outages, we write $o_1\le o_2\le ...\le o_n$ for the outage times in the order in which they occur and $r_1\le r_2\le ...\le r_n$ for the restore times in the order in which they occur.\footnote{Since the restores times $r_1\le r_2\le...\le r_n$ are in the order in which they occur, they are usually not in the same order as their corresponding outages. For example, $r_1$ is the restore time of the outage that gets restored first, and this may or may not be the restore of the outage $o_1$ that occurred first.}
The outages happen in the time interval $[o_1,o_n]$ and the restores happen in the time interval $[r_1,r_n]$.
In real data the restores typically start before the outages end, so these time intervals overlap.
We write $c_k$ for the number of customers outaged at the $k$th outage.

The component performance curve $P(t)$ tracking the number of unrestored outages easily generalizes to a customer performance curve $P^{\rm cust}(t)$ that tracks the number of unrestored customers: $P^{\rm cust}(t)$ is the negative of the cumulative number of unrestored customers in an event.

It is now straightforward to give formulas for the resilience metrics that we evaluate for each event:
\begin{itemize}
  \item {\sl event size} = number of outages $=n$
  \item {\sl outage hours} = area under performance curve\\ $=r_1-o_1+r_2-o_2+...+r_n-o_n=-\int_{o_1}^{r_n}P(t)dt$
  \item {\sl event duration} $=r_n-o_1$
  \item {\sl time to first restore} $=r_1-o_1$
  \item {\sl restore duration} $=r_n-r_1$
  \item {\sl restore rate} $=n/(r_n-r_1)$
  \item {\sl outage rate} $=n/(o_n-o_1)$
  \item {\sl customers out} $= c_1+c_2+...+c_n$
  \item {\sl customer hours} = area under customer performance curve $=c_1(r_1-o_1)+c_2(r_2-o_2)+...+c_n(r_n-o_n)$ $=-\int_{o_1}^{r_n}P^{\rm cust}(t)dt$
\end{itemize}
The two expressions for {\sl outage hours}, or for {\sl customer hours}, are shown to be equal in \cite{DobsonPESL23}.
Note that dividing the {\sl customer hours} for an event by the number of customers gives the contribution of that event to SAIDI, assuming for the larger events that major event days are included in SAIDI. 

\section{Outage sampling to get the average metrics for reduced outage rates}  \label{SamplingSection}
This section describes the sampling from the historical outages to select a reduced number of outages that represents hardening.
The resilience metrics are recalculated for many such samples and then averaged to find the average improvements in the metrics.
This metric calculation is applied separately to small, medium, and large events.

Suppose that the mean outage rate $\overline{F}(v)$ at wind speed $v$ is calculated from the $k$ outages $\left \{ e_{1},...,e_{k} \right \}$. 
According to section \ref{hardening}, a shift in the area outage rate curve gives the new outage rate $\overline{F}_{\rm new}(v)$ at wind speed $v$. 
To realize this reduced outage rate, we randomly sample $k_{new}$ outages from $\left \{ e_{1},...,e_{k} \right \}$ where, in general, 
\begin{align}
   k_{new} = \mbox{round}\bigg( k\frac{\overline{F}_{\rm new}(v)}{\overline{F}(v)}\bigg)
   \label{eq:OutagesSubsetSizeNew}
\end{align}
But in our case of exponential outage rate curves, (\ref{eq:OutagesSubsetSizeNew}) simplifies using (\ref{eq:exponentialModelShift}) to 
\begin{align}
   k_{new} = \mbox{round}( k e^{-bx})\label{eq:OutagesSubsetSizeNewexp}
\end{align}

At each wind speed, we sample a reduced number of outages from the historical outages to obtain a new set of outages that realizes the new area outage rate and the effect of the hardening.
For example, if there are 50 historical outages at 25 mph wind speed, and the hardening reduces the outage rate at 25 mph by 10\%, then we randomly sample 45 outages from the 50 outages. This sampling is done for all the wind speeds.
We then calculate the new resilience metrics for the new set of outages.
It is convenient to write $M$ for one of these resilience metrics, and $M_{\rm new}^{(1)}$ for the metric evaluated on the new set of outages at all wind speeds.
The entire sampling and metric evaluation procedure is then repeated $m$ times to obtain the new metrics $M_{\rm new}^{(1)}$, $M_{\rm new}^{(2)}$, ..., $M_{\rm new}^{(m)}$.
Finally, the average new metric is computed as 
\begin{align}
   \overline{M}_{\rm new} = \frac{1}{m}\sum_{i=1}^m M_{\rm new}^{(i)}\label{eq:AverageM}
\end{align}

An explanation for this procedure is that while the shift in the outage rate curve determines the new reduced number of outages $k_{\rm new}$ at each wind speed, it does not determine {\sl which} outages are to be omitted when realizing this reduced number of outages.
That is, we do not know which outages at each wind speed will be removed by the hardening.
Therefore, we compute the average new metric $\overline{M}_{\rm new}$ for random samples of the reduced number of outages. 

One complication is that the sampling can sometimes remove outages from an event in such a way that the event splits into smaller events.
This complication is handled with super events as explained in section \ref{superevents}.

For our calculations, the number of repetitions of the sampling procedure is chosen as $m=2000$ to ensure that the confidence interval $\overline{M}_{\rm new}\pm 0.01$ contains the true value of the mean metric with probability 99\% or greater.
$m=2000$ is obtained as follows: Since the distributions of the sampled metrics $M_{\rm new}$ are observed to be approximately normal, the half width $d$ of a $99\%$ confidence interval for the mean $\overline{M}_{\rm new}$ satisfies 
\begin{align}   
    d\leq t_{0.005,m-1}\frac{s}{\sqrt{m}}           \label{eq:standardError}
\end{align}
where $s$ is the sample standard deviation of the metric samples %$M_{\rm new}^{(1)}$, $M_{\rm new}^{(2)}$, ..., $M_{\rm new}^{(m)}$ 
and $t_{0.005,m-1}$ is the 99.5\% percentile of the Student-$t$ distribution with $m-1$ degrees of freedom. 
We take $d=0.01$ and increase $m$ until (\ref{eq:standardError}) is satisfied for each metric.

There is a clear pattern in the data of far more smaller events and much fewer large events. 
For example, Fig.~\ref{fig:eventSizePlotMB} shows the empirical probability distribution of {\sl event size} for area~1 on a log-log scale. 
This pattern affects the processing of the results because if one averages all the results together, the smaller events will dominate the average.
To address this, and particularly because resilience must have some focus on the large events, we divide the events into small events (1 or 2 outages), medium events (3 to 15 outages), and large events (16 or more outages). 
Area 1 has 2386 small events, 526 medium events, and 32 large events; area 2 has 2845 small events, 773 medium events, and 88 large events.
Average metrics for small, medium, or large events can distinguish the resilience performance for these different sizes of events, while still having enough large events to give usable estimates of the average metrics for large events.

\begin{figure}[htb]
    \centering
    \includegraphics[width=0.48\textwidth]{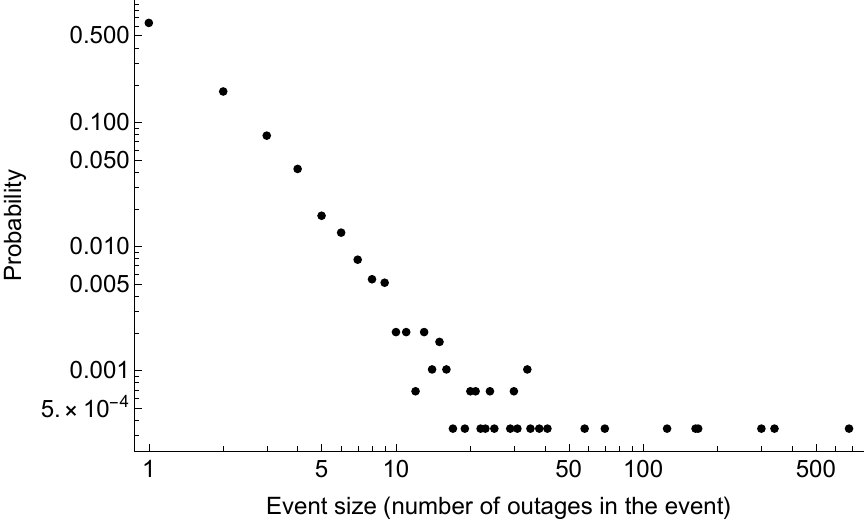}
    \caption{Empirical distribution of {\sl event size} in area 1 (log-log scale)}
    \label{fig:eventSizePlotMB}
\end{figure}

\section{Representing earlier or faster restoration}
\label{restoration}

%We have already discussed the impacts of 10\% wind hardening investments on the resilience metrics of the distribution system in previous sections. In this section, we discuss the impacts of investments to improve service restoration and compare them with the hardening investments.

This section represents the effects of improved restoration by modifying the restore times of the historical data.
Recall that in an event with $n$ outages, we write $r_1\le r_2\le ...\le r_n$ for the restore times in the order in which they occur. The outage times of the components that are restored in this restore order are written as $o_{\pi(1)}, o_{\pi(2)},...,o_{\pi(n)}$. The components do not usually outage in the order in which they are restored, so the $\pi$ function permutes the order to account for this.

We represent the improved restoration in two ways.
First, the repair can start earlier by providing more resources for identifying, locating, and automatically resolving faults; this includes investing in more sensors, switches, communications, meters, and reclosers, as well as more crews to inspect the lines and clear debris.
Let the change in start time be specified by $t_{\rm earlier}$, then the new restore times for the event are 
\begin{align}
    r_k^{\rm new}=\max\{r_k-t_{\rm earlier},o_{\pi(k)}\}, \quad k=1,...,n.
    \label{rkearlier}
\end{align}
Taking the maximum with $o_{\pi(k)}$ in (\ref{rkearlier}) limits the new restore time so that $r_k^{\rm new}\ge o_{\pi(k)}$; restoration of a component must occur after its outage.
%Limiting the shortening time to the component repair time $\rho_k$ ensures that the component repair times remain nonnegative.

Second, the rate of restoration can be increased and the restoration duration decreased by investing in more repair crews, better stocks of spare parts, and better route scheduling.
Let the faster restore duration be specified by multiplying by a factor $c_{\rm faster}<1$.
Then the restore duration of the $k$th restore $r_k-r_1$ is reduced by a factor of $c_{\rm faster}$, as long as the new restore time occurs after its corresponding outage:
\begin{align}
    r_k^{\rm new}=\max\{r_1+(r_k-r_1)c_{\rm faster},o_{\pi(k)}\}, \  k=1,...,n.
    \label{eq:rnewrate}
\end{align}
%Note that in the common case of the new restore times not being limited by the component restore times, this investment decreases the restore duration $r_n-r_1$ by a factor of $d$ and increases the average restore rate $n/(r_n-r_1)$ by a factor $1/d$.

\section{Results}
\label{results}

\begin{table*}[!t]
    \renewcommand{\arraystretch}{1.1}
    \caption{Average metrics for small, medium, and large events and their changes with hardening or earlier or faster restoration}
    \label{tab:resultsArea1}
    \centering
    \begin{tabular}{l c c c c c c c c c c c c}
    \multicolumn{13}{c}{AREA 1}\\
        \hline
&\multicolumn{3}{c}{Base Case Events}&\multicolumn{3}{c}{Change with Hardening}&\multicolumn{3}{c}{Change with Earlier Restore}&\multicolumn{3}{c}{Change with Faster Restore}\\[-2pt]
Resilience Metric&small&medium&large&small&medium&large&small&medium&large&small&medium&large\\
        \hline
        {\sl event size}              & 1.22    & 4.64    & 78.91   & -10.0\%    & -10.0\%   & -10.0\%   & 0\%     & 0\%        & 0\%      & 0\%      & 0\%      & 0\%      \\
        {\sl outage hours}            & 3.06    & 17.33   & 2084    & -10.0\%    & -10.0\%   & -10.0\%   & -74.8\% & -58.6\%    & -10.0\%  & -0.6\%   & -5.7\%   & -10.0\%  \\
        {\sl event duration}          & 2.72    & 8.22    & 46.62   & -8.9\%     & -4.5\%    & -1.2\%    & -68.2\% & -27.5\%    & -5.5\%   & -0.7\%   & -3.9\%   & -4.4\%  \\
        {\sl time to first restore}   & 2.40    & 2.97    & 2.95    & -7.6\%     & 1.7\%     & 2.2\%     & -75.9\% & -65.1\%    & -69.0\%  & 0\%      & 0\%      & 0\%      \\
        {\sl restore duration}        & 0.32    & 5.25    & 43.67   & -19.0\%    & -10.7\%   & -2.4\%    & -10.0\% & -6.3\%     & -1.2\%   & -6.2\%   & -6.1\%   & -4.7\%   \\
        {\sl customers out}           & 45.26   & 257.93  & 5610    & -10.0\%     & -10.0\%   & -9.9\%   & 0\%     & 0\%        & 0\%      & 0\%      & 0\%      & 0\%      \\
        {\sl customer hours}          & 85.88   & 680.83  & 95304   & -10.0\%     & -10.1\%    & -10.0\%   & -80.5\% & -68.2\%    & -14.1\%  & -0.5\%   & -5.4\%   & -9.1\%   \\[2pt]
        \multicolumn{13}{c}{AREA 2}\\
        \hline
        {\sl event size}              & 1.25    & 4.75    & 62.48   & -10.0\%    & -10.0\%   & -10.0\%   & 0\%     & 0\%        & 0\%      & 0\%      & 0\%      & 0\%      \\
        {\sl outage hours}            & 2.71    & 15.28   & 1142    & -10.0\%    & -10.0\%   & -10.0\%   & -67.8\% & -51.1\%    & -10.0\%  & -0.5\%   & -4.6\%   & -10.0\%  \\
        {\sl event duration}          & 2.39    & 7.51    & 37.67   & -8.8\%     & -4.5\%    & -1.3\%    & -61.7\% & -22.7\%    & -4.8\%   & -0.6\%   & -3.1\%   & -4.0\%   \\
        {\sl time to first restore}   & 2.10    & 2.66    & 2.56    & -7.4\%     & 1.6\%     & 3.0\%     & -68.6\% & -56.8\%    & -62.3\%  & 0\%      & 0\%      & 0\%      \\
        {\sl restore duration}        & 0.28    & 4.85    & 35.11   & -19.0\%    & -10.8\%   & -2.9\%    & -10.5\% & -3.9\%     & -0.6\%   & -4.8\%   & -4.8\%   & -4.3\%   \\
        {\sl customers out}           & 50.74   & 242.98  & 4084    & -10.0\%    & -10.1\%   & -10.0\%   & 0\%     & 0\%        & 0\%      & 0\%      & 0\%      & 0\%      \\
        {\sl customer hours}          & 85.51   & 547.28  & 58700   & -10.0\%    & -10.0\%   & -10.1\%   & -74.0\% & -61.9\%    & -12.0\%  & -0.3\%   & -4.7\%   & -9.0\%   \\
        \hline
        &\multicolumn{3}{c}{all time quantities in hours}&\multicolumn{9}{c}{small events have 1--2 outages, medium events have 3--15 outages, large events have $\ge$16 outages}\\
    \end{tabular}
\end{table*}

This section presents a case study of the impacts of hardening and improved restorations on the resilience metrics for areas 1 and 2 of the distribution system. 

\subsection{Base case resilience metrics}
The base case is the historical outages without any modifications.
Table \ref{tab:resultsArea1} shows the base case average values of the resilience metrics for small events (1 or 2 outages), medium events (3--15 outages), and large events ($\ge$16 outages).
Considering the different sizes of events separately and with some special attention to the large events is needed for this quantification of resilience, as explained at the end of section~\ref{SamplingSection}. 
As expected, all the metrics (except {\sl time to first restore}) clearly show the increased impact on customers as the event size increases from small to medium to large. 
The average resilience metrics show that area 1 has greater customer impacts than area 2 for large events.

\subsection{Change in metrics due to hardening}
The hardening for each area with respect to wind is represented by an increased mile-per-hour wind rating, which gives a percentage reduction in the outage rate (see section \ref{hardening}) that is implemented by sampling a reduced number of outages (see section \ref{SamplingSection}). 
For the case study, it is convenient to consider a hardening that gives a 10\% reduction in the outage rate for both areas.
This 10\% reduction in outage rate corresponds to 0.28 mph wind hardening for area 1 and 0.22 mph wind hardening for area 2.
The 10\% reduction in outage rate is implemented by sampling 10\% fewer outages, so that the hardening reduces the average {\sl event size} (number of outages) by exactly 10\%, as confirmed in Table \ref{tab:resultsArea1}.
The hardening also reduces the average {\sl outage hours} by exactly 10\%. This result follows from a resilience metric formula in \cite{DobsonPESL23}.\footnote{
For each event, \cite[(17)]{DobsonPESL23} gives {\sl outage hours}
$=n \overline{\rho}=n \left(\frac{1}{n}\sum_{k=1}^n \rho_k\right)$,
 which is the number of outages $n$ times the average component restore time $\overline{\rho}$ in the event. 
 Averaging over all the samples with $n$ reduced by 10\% leaves the expectation of the average of $\overline{\rho}$ over all the samples equal to $\overline{\rho}$. Therefore the average {\sl outage hours} reduces by exactly 10\%.}

Table \ref{tab:resultsArea1} shows that the hardening decreases the {\sl customers out} and {\sl customer hours} by approximately 10\% for events of all sizes in both areas. 
However, the reductions in the average duration metrics are less than 10\%, except for the {\sl restore duration} of small events.
When we sample the reduced number of outages, one or more outages get removed from events randomly.
Depending on exactly which outage is removed, the {\sl event duration} and {\sl restore duration} metrics either decrease or remain the same.
The {\sl restore duration} of small-size events reduces by more than 10\% because the {\sl restore duration} drops to zero when an outage is randomly removed for an event having only two outages.
Removing an outage has a larger percentage impact on durations in a small event than in a large event, so the average decrease is smaller for large events. 
In particular, it is notable that the hardening which gives close to a 10\% average reduction in {\sl customer hours} reduces the average {\sl event duration} for large events by less than 2\%.

\looseness=-1
Table \ref{tab:resultsArea1} shows that {\sl time to first restore} can increase or decrease when outages are removed from an event. This is because {\sl time to first restore} is the time difference between the first outage and the first restored outage. 
For example, if the sampling retains the first outage but removes the first restored outage, the {\rm time to first restore} can increase. 
If the sampling removes the first outage but retains the first restored outage, the {\sl time to first restore} can decrease.
And if the first outage is the same outage as the first restored outage, and the sampling removes that outage, then {\sl time to first restore} can increase or decrease.
{\sl Time to first restore} is the only metric that can increase after sampling removes outages; all the other metrics can only decrease or remain the same when outages are removed. 
A detailed discussion along with the probability of change in each metric due to outages removed by sampling is given in \cite{AhmadMS23}.

\subsection{Change in metrics due to earlier or faster restoration}
Table \ref{tab:resultsArea1} also presents the change in metrics for the two types of improved restoration.
The earlier restoration time is specified by the shortening time $t_{\rm earlier}$ in (\ref{rkearlier}), and 
the faster restoration is specified by the improvement factor $c_{\rm faster}$ in (\ref{eq:rnewrate}). 
It is convenient in order to facilitate comparisons to select $t_{\rm earlier}$ and $c_{\rm faster}$ for each area so that the average {\sl outage hours} for large events decrease by exactly 10\%.
In particular, we select $t_{\rm earlier}=2.84$ hour and $c_{\rm faster}=0.9385$ for area 1 and $t_{\rm earlier}=1.92$ hour and $c_{\rm faster}=0.9522$ for area 2.

The earlier and faster restorations do not remove any outages, so the {\sl event size} and {\sl customers out} metrics stay the same.

%in the case of earlier restorations and the improvement factor $c_{\rm faster}$ in the case of faster restorations are selected to have an exact 10\% decrease in the outage-hours of events in the large size category. 
%This choice is made to make the changes due to improved restorations comparable to those due to hardening. 
%Moreover, since we are more interested in studying the impacts of large-size events, the values are selected for large events. The same values are applied to the small and medium size events categories to see the impacts.

%In the case of earlier restorations, a shortening time of 115.2 minutes and 170.2 minutes gives a 10\% reduction in the mean outage-hours of large events of areas 1 and 2, respectively. 
%Therefore if investments were made such that the restoration process could be started earlier by these values, then we would have seen a 10\% decrease, on average, in the outage-hours of large-size events in both areas. 

The earlier restoration in (\ref{rkearlier}) reduces restoration times by $t_{\rm earlier}$, but with a limitation so that restoration of each component does not occur before it outages. 
This limitation for some restores causes the average {\sl time to first restore} to reduce but by less than $t_{\rm earlier}$.
For large events, the average {\sl time to first restore} reduces from 2.95 hrs to 0.92 hrs and from 2.56 hrs to 0.96 hrs in area 1 and area 2, respectively. 
%It is because we can only start the restore process as soon as the outage process starts and cannot start it before that, and therefore for some of the events, we cannot shorten the time exactly by $t_{\rm earlier}$.
In the absence of the limitation in (\ref{rkearlier}), all restoration times are reduced by $t_{\rm earlier}$ and the {\sl restore duration} stays the same.
However, the limitation affects some of these restorations, which can include the first and last restoration of an event, resulting in the average {\sl restore duration} decreasing.
%Similarly, the change in restore duration should be zero as, in the ideal case, we only shift the restore process to an earlier time, keeping the restore duration the same. However, we can only shift the restore process by an amount equal to the duration (repair time) of the last restored outage $r_n$, and also, we cannot shift it more than the duration of the first restored outage $r_1$. Therefore the restore duration either increases or decreases depending on $t_{\rm earlier}\lesseqgtr r_n$ and $t_{\rm earlier}\lesseqgtr r_1$. An overall decrease in the mean restore duration of both areas shows that the shift gets limited by $r_1$ more often on average.

%For the faster restoration case, an improvement factor of 0.9522 and 0.9385 gives a 10\% reduction in the mean outage-hours of large events of areas 1 and 2, respectively. In other words, if investments were made to increase the restore rates by approximately 5\% and 6.6\%, then those would have resulted in a 10\% reduction in the mean outage-hours of large-size events of area 1 and area 2, respectively. 

The faster restoration in (\ref{eq:rnewrate}) speeds up the restoration by the factor $c_{\rm faster}$ over the duration of the restoration, but with a limitation so that restoration of each
component does not occur before it outages. 
In the absence of the limitation in (\ref{eq:rnewrate}), the {\sl restore duration} would decrease by 6.2\% and 4.8\% corresponding to the $c_{\rm faster}$ values for area 1 and area 2, but we see in Table \ref{tab:resultsArea1} less decrease in some of the mean {\sl restore durations} due to the limitation.
The {\sl time to the first restore} is unaffected by the faster restoration. 
The change in the average {\sl customer hours} quite closely follows the change in the average {\sl outage hours}. 
The faster restoration affects the small events differently because they have only one or two outages.
The events with only one outage remain unchanged as their {\sl restore duration} is zero.
Also, a large number of two-outage events have zero {\sl restore duration} when both outages are restored at the same time.
Therefore, those small events also remain unchanged. Consequently, only a small proportion of small events see improvements due to faster restoration, which explains the very small changes in the average metrics other than {\sl restore duration}.

\subsection{Comparing hardening and improved restoration}
Two overall options are available for power system resilience investments. One is to invest in hardening and the other is to invest in improved restoration.
Hardening invests in infrastructure, whereas improved restoration invests in crews, their resources, and automated actions.
Our results confirm the general observation that hardening reduces the {\sl event size} (number of outages) while affecting event durations less, whereas improved restorations decrease the outage durations but do not affect the {\sl event size}. 

The results in table \ref{tab:resultsArea1} show that a hardening decreasing {\sl outage hours} and {\sl event size} by 10\% also decreases {\sl customer hours} and {\sl customers out} by approximately 10\%, irrespective of small, medium or large events.
The {\sl customer hours} are particularly important in assessing the impact of power outages on consumers.
On the other hand, the improved restorations to achieve a 10\% reduction in the mean {\sl outage hours} for large events also provide almost the same percentage decrease in {\sl customer hours}. However, unlike hardening, the {\sl event size} and {\sl customers out} remain unchanged.
So the customers would still face power outages but get restored more quickly.

All these results show how different overall investments would have changed the various resilience metrics and customer impacts. This quantifies the various benefits that the investments would have made to the utility and its customers.

\section{Sampling and super events}
\label{superevents}

Since the sampling removes outages, the remaining outages that were in the same event before sampling may not all be overlapping after sampling and so can sometimes split into two or more events. 
We call the set of events arising after sampling in this way from one event before sampling a ``super event", and this section explains super events.
For example, consider the timeline plot of a typical event $E=\{e_1$, $e_2$, $e_3$, $e_4$, $e_5$, $e_6$, $e_7$, $e_8$, $e_9\}$ with 9 outages before sampling as shown in Fig.~\ref{fig:resilienceEventTimeline}.  
If the sampling removes $e_1$, then the remaining outages still form one event and the super event is $\{\{e_2$, $e_3$, $e_4$, $e_5$, $e_6$, $e_7$, $e_8$, $e_9\}\}$. 
However, if the sampling removes $e_4$ and $e_5$ then the remaining outages form two events since the system is fully restored when outage $e_3$ is restored and the super event is $\{\{e_1, e_2, e_3\}, \{e_6, e_7, e_8, e_9\}\}$. 

The changes in the sizes and numbers of events due to sampling cause problems when the metrics of events before and after sampling are compared: 
Basic to the analysis is the classification of events by size based on their number of outages, and the variable reduction in event size, and especially events splitting into multiple smaller events, interferes with tracking the effect of the sampling on event metrics and disrupts the effect of the sampling on the categories of small, medium, and large events, since the events can change categories after sampling. 
These problems are resolved by keeping track of all the events arising after sampling from one event before sampling in a super event, and appropriately defining the metric of a super event as follows:

Consider a metric $M$ that can be evaluated on an event $E$ as $M[E]$, and a super event $\{E_1,E_2,...,E_p\}$ that has events $E_1,E_2,...,E_p$ after sampling. Then, for the metrics 
{\sl event size} (number of outages),
{\sl outage hours},
{\sl restore duration},
{\sl number of customers out}, and
{\sl customer hours}, we define the metric evaluated on the super event as 
\begin{align}
    M[\{E_1,E_2,...,E_p\}]=
    M[E_1]+M[E_2]+...+M[E_p]\notag
\end{align}
and for the metrics 
{\sl restore rate},
{\sl outage rate}, and
{\sl time to first restore}, we replace summation by the average to define
\begin{align}
    M[\{E_1,E_2,...,E_p\}]=\mbox{$\frac{1}{p}$}
    (M[E_1]+M[E_2]+...+M[E_p])\notag
\end{align}
Events with only one outage can disappear if that outage is removed by sampling. In this case, the super event is the empty set $\{\,\}$ and all the metrics evaluate to zero.

\begin{figure}[htb]
    \centering
    \includegraphics[width=0.48\textwidth]{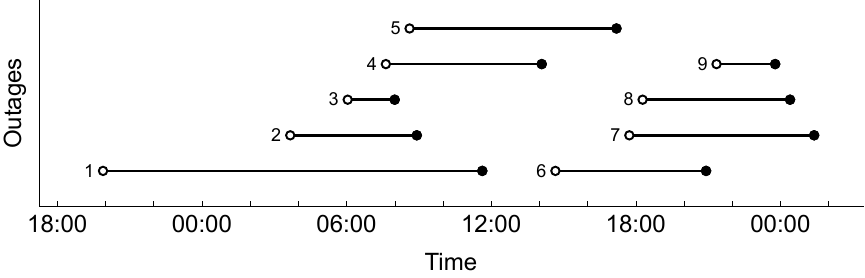}
    \caption{Timeline plot of a medium-sized resilience event with 9 overlapping outages shown by line segments. For each line segment, the open circle is the outage start time and the dot is the outage restore time.}
    \label{fig:resilienceEventTimeline}
\end{figure}

\section{Comparing historical data and model-based approaches}
The new approach towards quantifying wind resilience investment based on ``rerunning history" with the changes to outages and restores caused by the investment can be compared with approaches based on models or models and data to predict the future effect of the investments. 
This is necessarily a preliminary overall comparison, because while hundreds of papers have explored aspects of prediction with models, this seems to be the first paper outlining the historical approach, and so its full potential and limitations are not yet explored. 
Moreover, many of the model-based approaches have legitimately different objectives than this paper, including modeling only some of the processes, such as modeling only the weather impact on the power grid, modeling and optimizing only the restoration process, analyzing specific hurricanes, or optimizing an operational response to a specific severe weather forecast. There is no strict dividing line between model-based and data approaches because machine learning models are data-intensive, and physically based models use data for parameter values, calibration, and validation.

Weather shows considerable variability in space and time. 
A historical approach assumes the weather that previously happened, which has similarities to future weather but is not the same as future weather, whereas predicting the future weather can look at specific likely cases, but there is substantial uncertainty in its predictions. 
The historical approach records the actual response of the power system to the experienced weather, and it is feasible to estimate the overall changes in outages and restores that would have been caused by the investment. 
The historical approach accounts for combined effect of all the detailed differences in the individual power system components and their individual weather conditions, as well as all the changes (both planned and unexpected) in the power system and its operation and maintenance over the historical period. 
A model-based approach must make many assumptions approximating the power system and its response to weather. 
The representative physical models of power system components used in the model-based approach to estimate component fragility can be calibrated and validated with data. 
Given specific simulated outages and a standard detailed specification of the power system including its protection, it is relatively straightforward to compute the customers disconnected. 
The restoration is harder to model, but it can be done. 
However, there is uncertainty in the simulated outages due to the uncertainties in the weather, and in realizing the fragility curves, and in restoration so that the simulation of all these processes must be run many times to produce samples of the outages and restores.
That is, the model-based approach is computationally expensive, whereas the computations for the historically-based method are easy and quick.
The computation of resilience metrics from the historical or simulated events is straightforward in both approaches.

The model-based approach has the advantage that the proposed investment can be more directly expressed in the model, whereas the historical data approach needs to estimate the effect of the investment on outages and restores. 
Detailed engineering is needed to make this estimate, but utilities do this sort of detailed engineering routinely, and the estimate can be made using engineering experience, data from previous investments, or a model. 

\looseness=-1
The historical data approach has a specific, realistic, and easily understood assumption of the past history. 
That is, the historical data approach is based on how the power system, with all of its complexity and changes, actually responded to the wind stress that occurred and then evaluates the effects that the investment would have had.  
%The historical data approach requires an estimate of the impact of the specific investment on outages and restores.
Building and running validated models for weather and power system outages and restoration requires far greater effort, but the investments can be expressed in the model more directly and more flexibly. The models can choose their assumptions and approximations about the future, but have to handle the uncertainties of prediction.
Justifying a proposed investment to stakeholders by looking back at what its effect would have been and by predicting its future effects are clearly complementary, and both can be pursued.

\section{Conclusions}
\label{conclusions}

\looseness=-1
This paper combines historical outage and weather data to construct area outage rate curves to quantify the resilience of areas of a distribution system to wind. 
An investment hardening the distribution system would have shifted the area outage rate curves and reduced the outage rates, and the effect of this on resilience metrics is quantified by sampling a reduced number of the historical outages and recalculating the resilience metrics. 
The effect of an improvement in restoration times is also quantified by advancing or speeding up the historical restoration and recalculating the metrics.
The resilience metrics include the {\sl event size} (number of outages), durations, rates, and {\sl customer hours} evaluated on resilience events of different sizes. 
These data-driven calculations quantify the impact on customers that previous investments would have had.

Overall, we initiate a new approach towards resilience quantification.
Specific contributions and attributes of this new approach are:
\begin{itemize}
    \item Quantifying the impact on customers that a resilience investment would have made in the past gives a novel and credible way to justify the benefits of the investment that can be tangible to utilities, communities, and regulators, because it clearly shows how the lived past experience of customers would have been improved.
    This is significant since effective ways to justify resilience investments to stakeholders are essential for practically implementing resilience. Quantifying the effect that the investment would have made in the past complements and augments justifications for resilience investments that rely on projections into the future with models. 
    \item Modifying historical data is an entirely new way to quantify resilience and resilience investments that calculates the changes in standard resilience metrics from the effects that the investments would have had. This approach directly driven by data has clear advantages in realism in accounting for all the conditions that the power system experienced over the period of observation, including variations in space and time in weather, load, upgrades, operating procedures and restoration policies, and component design, location, conditions and maintenance. No modeling assumptions are made. 
    \item We construct area outage rate curves that quantify the wind resilience of an area of a distribution system directly from data by describing how the mean outage rate of the area increases as a specific nearby wind measurement increases. The area outage rate curves have a similar form as component fragility curves, but describe the empirical aggregate area response in terms of outages rather than the response to wind experienced at specific components of the distribution system. 
    Area outage rate curves rework the concept of fragility curves for a different purpose.
%     Area outage rate curves are not suitable to be directly inserted into a component fragility model.
    \item The overall effect of resilience investments are simply represented by an earlier or faster restoration or by a hardening that increases resilience to wind by a given number of miles per hour. This enables a novel and credible comparison of investment in hardening versus investment in better restoration in terms of customer impact. The hardening is implemented on the historical outages in a novel way: 
    The outage rate curve is shifted by the hardening to determine the reduced number of outages at each wind speed. Then the reduced number of the outages are sampled from the historical outages and the improved metrics are calculated. 
    Repeating this many times and averaging gives the expected improvement in the metrics. The effect of the improved restoration is implemented by changing the historical restoration times to start earlier or to be completed faster.
    \item %Wind hazard hardening by a given number of miles per hour shift area outage rate curves and reduce outage rates. The effect that the reduced outage rates would have had are implemented by sampling subsets of outages from the historical data. Resilience metrics of small, medium, and large events are calculated from the sampled outages and compared with the original resilience metrics to assess the impact of the hardening investments. 
    The technical aspects of the calculations include: \\(a) the segregation of resilience events into small, medium, and large events to get a meaningful assessment of the resilience of events of different sizes with metrics, (b) the new concept of super events to track the metrics of events that split into smaller events when outages are removed, and \\ (c) leveraging recent work \cite{CarringtonPS21} that automatically extracts events of all sizes from utility outage data and calculates a range of metrics for each event. 
    %The calculated impact of the hardening or faster restoration can be analytically confirmed for some metrics.
    \item The method is limited to a net reduction in historical outages; it does not synthesize additional new outages. Note that this limitation does not rule out the incorporation of all future effects. For example, the effect of a future increase in the average wind speed can still be represented, as long as it is offset by sufficient hardening so that the net outage rate decreases. 
    Also the effect of a percentage increase in the rate of events is straightforward to evaluate because the metrics per event are unchanged, and any metrics that accumulate over a time period increase by the same percentage. The feasibility of these extensions is significant since wind speeds and storm frequencies are expected to increase with climate change.
    \item Many standard definitions of resilience, such as \cite{CIGREELECTRA19}, focus exclusively on high-impact, low-frequency events. However, consider the following reason to extend the scope of methods for quantifying resilience investments to all sizes of events: A resilience investment will generally change the probability of events of all sizes, and to properly assess the benefits of the investment we need to quantify its benefits for all event sizes. One simple way to do this and easily communicate the results divides the events into small, medium, and large events, and assesses the change in resilience metrics for the small events, the medium events, and the large events.
    \item Investments in the power system result in specific hardening and/or restoration improvements in specific parts of the system. This paper models the overall effect of improvements by changes in outages and restores and then quantifies the impacts on resilience metrics of these overall improvements. However, description of the specific hardening/restoration improvements, their cost, and estimation of the change in outages and restores from the specific projects are not done in this paper due to the unavailability of the relevant feeder and cost data. We hope to find suitable outage, feeder, weather, and cost information for a different distribution system so that we can address this aspect in future work. We note that utilities already have extensive experience in proposing, costing, and estimating the impact of projects.
    \item The outage data and weather data required to use this approach are easily available to any distribution utility with an outage management system, and the computations are relatively straightforward and fast.
\end{itemize}

There is promising scope for extending the methods of this paper in future work: Other wind data could be tested and evaluated.
Other hazards such as flood or icing could be considered and outage cause code information could be related to the hazards and leveraged.
Detailed model-based approaches could be used to directly link specific engineering improvements to the overall changes in hardening and faster restoration that model the investments in this paper. 
If cost estimates and probability estimates of rarer events can be improved, then the risk and improvements to risk could be quantified.

This paper has established the feasibility and useful characteristics of a new approach to quantifying resilience and the benefits of investments in resilience from historical data, and we are confident that further developments can follow.

\renewcommand*{\UrlFont}{\rmfamily}
\renewcommand*{\bibfont}{\footnotesize}
\printbibliography %Prints bibliography

\begin{IEEEbiographynophoto}
{Arslan Ahmad}
(Student Member, IEEE) received a B.S. degree in electrical engineering from Punjab University, Pakistan in 2013 and an M.S. degree in electrical engineering from Iowa State University, USA in 2023.
He has over five years (2016-2021) of experience in power distribution system operations. He is a recipient of the Fulbright Scholarship (2021-2023) and the University of Twente Scholarship (2015-2016).
He is pursuing a Ph.D. in electrical engineering at Iowa State University,  USA. His research interests are data analytics and applications of statistical and ML methods to improve the resilience and reliability of power systems.
\end{IEEEbiographynophoto}

\begin{IEEEbiographynophoto}
{Ian Dobson}
(Fellow, IEEE) received the B.A. degree in mathematics from Cambridge University, UK, and the Ph.D. degree in electrical engineering from Cornell University, USA. He is currently Sandbulte Professor of Engineering at Iowa State University, Ames, IA, USA. His interests are blackout risk and  power system resilience and stability. 
\end{IEEEbiographynophoto}

\end{document}